\documentclass[12pt]{article}
\usepackage{amsmath}
\def\be{\begin{equation}}
\def\ee{\end{equation}}
\def\arr{\begin{array}{rll}}
\def\ea{\end{array}}
\def\bea{\begin{eqnarray}}
\def\eea{\end{eqnarray}}

\def\N2{$N{=}2$}

\def\>{\rangle}
\def\<{\langle}
\def\+{\dagger}
\def\={\ =\ }

\begin{document}
\renewcommand{\thefootnote}{\fnsymbol{footnote}}
\begin{titlepage}
\setcounter{page}{0}
\vskip 1cm
\begin{center}
{\LARGE\bf Casimir operators of centrally extended }\\
\vskip 0.5cm
{\LARGE\bf  $l$--conformal Galilei algebra}\\
\vskip 1.5cm
$
\textrm{\Large Anton Galajinsky\ }^{a}, \quad
\textrm{\Large Ivan Masterov\ }^{b}
$
\vskip 0.7cm
${}^{a}$ {\it
Tomsk Polytechnic University, 634050 Tomsk, Lenin Ave. 30, Russia} \\
\vskip 0.2cm
${}^{b}$ {\it
Tomsk State University of Control Systems and Radioelectronics, 634050 Tomsk, Lenin Ave. 40, Russia} \\
\vskip 0.2cm
{Emails: galajin@tpu.ru, iv.masterov@yandex.ru}
\vskip 0.5cm

\end{center}
\vskip 1cm
\begin{abstract} \noindent
The full set of Casimir elements of the centrally extended $l$--conformal Galilei algebra is found in simple and tractable form.
\end{abstract}


\vskip 0.5cm


\end{titlepage}

\renewcommand{\thefootnote}{\arabic{footnote}}
\setcounter{footnote}0

\noindent
Recently there has been a burst of activity in studying dynamical realizations of the $l$--conformal Galilei algebra \cite{LSZ}--\cite{KLS}. It is expected that such systems may prove useful for a deeper understanding of the non--relativistic holography. The term "$l$--conformal" stems from the fact that both the dimension of the algebra and its structure relations depend on an external parameter $l$, which may take integer or half--integer values \cite{NOR}. One more free parameter, which characterizes the algebra, is linked to the spatial dimension $d$.

In general, models with $l>\frac 12$ involve higher derivative terms. Making no harm to a free theory, such terms typically entail unstable classical dynamics or trouble with ghosts in quantum description as soon as the interaction is present. Currently known second order systems invariant under the $l$--conformal Galilei group include a set of the isotropic oscillators coupled to the external conformal mode \cite{FIL,GM} and the geodesic equations associated with a specific Ricci--flat metric of the ultrahyperbolic signature \cite{CG}.

Thus far, attention was mostly drawn to classical models. Potential quantum--mechanical applications pose the question of describing the Casimir operators of the centrally extended $l$--conformal Galilei algebra in simple and tractable form. In a very recent work \cite{AIM1}, a quartic Casimir invariant was constructed for the cases of $d=1$ and half--integer $l$, and $d=2$ and integer $l$. Worth mentioning also is the instance of $d=3$ and half--integer $l$ analyzed in \cite{AGM}. The goal of this brief note is to present the full set of the Casimir elements of the centrally extended $l$--conformal Galilei algebra for arbitrary values of $l$ and $d$.

Let us choose the basis, in which the structure relations of the algebra read
\begin{align}
&
[L_n,L_m] = (m - n)L_{n+m}, && [L_n,C_i^{(\alpha)}] = (\alpha-ln)C_i^{(n+\alpha)},
\nonumber\\[2pt]
&
[M_{ij},C_k^{(\alpha)}] = \delta_{ik}C_j^{(\alpha)} - \delta_{jk}C_i^{(\alpha)}, &&
[M_{ij},M_{kl}]=\delta_{ik} M_{jl}+\delta_{jl} M_{ik} - \delta_{il} M_{jk}-\delta_{jk} M_{il}.
\end{align}
Here $(L_{-1},L_0,L_1)$ designate the generators of temporal translations, dilatations, and special conformal transformations, respectively, $M_{ij}$, $i,j=1,\dots,d$, are associated with spatial rotations, and  $C^{(\alpha)}_i$, $\alpha=-l,\dots,l$, $i=1,\dots,d$, signify vector generators in the algebra. In this notation, $C^{(-l)}_i$ and $C^{(-l+1)}_i$ are linked to spatial translations, and Galilei boosts, while higher values of the index $\alpha$ correspond to accelerations.

In general, the algebra admits an extension by the central element $Z$ \cite{GM1}
\bea\label{cc}
[C_i^{(\alpha)},C_j^{(\beta)}]=(-1)^{\alpha+l}(\alpha+l)!(\beta+l)!Z\delta_{\alpha+\beta,0}\eta_{ij},
\eea
where $\eta_{ij}=\delta_{ij}$ for half-integer $l$ and arbitrary $d$, while $\eta_{ij}=\epsilon_{ij}$, $\epsilon_{ij}=-\epsilon_{ji}$, $\epsilon_{12}=1$, for integer $l$ and $d=2$. Keeping the central charge apart, the dimension of the algebra is equal to $3+\frac{d (d + 1)}{2}+2dl$.

Proceeding to the construction of the Casimir invariants, we first use the previous formula so as to disentangle $L_n$ from $C^{(\alpha)}_i$. Introducing the operators
\bea
\mathcal{L}_n = Z L_n - \frac{1}{2}\sum_{\alpha=-l}^{l}\frac{(-1)^{l-\alpha}(\alpha - ln)}{(l+\alpha)!(l-\alpha)!}
 C_i^{(-\alpha)}C_j^{(\alpha+n)} \eta_{ij},
\eea
one finds the relations
\begin{align}\label{strr}
&
[\mathcal{L}_n,C_i^{(\alpha)}]=0, && [\mathcal{L}_n,M_{ij}]=0,
\nonumber\\[2pt]
&
[\mathcal{L}_n,\mathcal{L}_m]=Z(m-n)\mathcal{L}_{n+m}, && [L_n,\mathcal{L}_m]=(m-n)\mathcal{L}_{n+m},
\end{align}
which all together imply that the operator
\bea\label{qc}
\frac{1}{2}\left(\mathcal{L}_{-1}\mathcal{L}_{1} + \mathcal{L}_{1}\mathcal{L}_{-1}\right) - \mathcal{L}_{0}^{2}
\eea
is a quartic Casimir invariant of the centrally extended $l$--conformal Galilei algebra. In particular, (\ref{qc}) encompasses all the cases previously studied in the literature \cite{AGM,AIM1}. When verifying the structure relations (\ref{strr}), the identity
\be
(\alpha-l n)-\frac{{(-1)}^n (\alpha+n(l+1))(l+\alpha)! (l-\alpha)! }{(l+n+\alpha)!(l-n-\alpha)!}=0
\ee
proves to be helpful.\footnote{When computing the Casimir invariants, the instances of $\alpha=l$, $n=1$ and $\alpha=-l$, $n=-1$ are naturally excluded from the sums at hand.}

In a similar fashion one can disentangle $M_{ij}$ from $C^{(\alpha)}_i$ by introducing the operators
\be
\mathcal{M}_{ij}=Z M_{ij}-\frac 12 \sum_{\alpha=-l}^{l}\frac{(-1)^{l-\alpha}}{(l+\alpha)!(l-\alpha)!}
\left( C_i^{(-\alpha)} \eta_{jk}C_k^{(\alpha)}-C_j^{(-\alpha)} \eta_{ik} C_k^{(\alpha)}\right),
\ee
which obey
\bea
&&
[C_i^{(\alpha)},\mathcal{M}_{jk}]=0, \qquad \qquad \qquad \qquad  [L_n,\mathcal{M}_{ij}]=0,
\nonumber\\[2pt]
&&
[M_{ij}, \mathcal{M}_{kl}]=\delta_{ik} \mathcal{M}_{jl}+\delta_{jl} \mathcal{M}_{ik} - \delta_{il} \mathcal{M}_{jk}-\delta_{jk} \mathcal{M}_{il}.
\eea
These allow one to build the Casimir elements associated with the $so(d)$ subalgebra. For odd $d$, there are $\frac{d-1}{2}$ such operators, which are obtained by the consecutive contraction of an even number of $\mathcal{M}$ with each other (see, e.g., Sect. 5 in \cite{FI}). For even $d$, one reveals $\frac{d}{2}$ invariants, which are constructed in a similar fashion, as well as one extra element resulting from the contraction of $\mathcal{M}$ with the Levi--Civit\'a totally antisymmetric tensor $\epsilon_{i_1\dots i_d}$.

Finally, according to the analysis in \cite{BB}, the total number of the Casimir elements characterizing a Lie algebra is given by the difference of its dimension and the rank of its commutator table regarded as a numerical matrix. It is straightforward to compute this number for the case at hand and verify that no further invariants ought to be expected. Thus, the centrally extended $l$--conformal Galilei algebra admits $\frac{d+1}{2}$ Casimir invariants if the dimension is odd and $\frac{d+2}{2}$ such operators if $d$ is even.

Concluding this work, we note that a similar analysis for the case of the vanishing central charge turns out to be surprisingly complex. The invariants constructed above, continue to hold after implementing the limit $Z\to 0$. Yet, a close inspection of various particular cases shows that, according to the criterion in \cite{BB}, extra Casimir elements ought to be expected, their number growing rather fast with $l$ and $d$ increasing. At the moment we lack a clear guiding principle of how to systematically build the missing invariants and hope to report on the progress elsewhere.

\vspace{0.5cm}

\noindent{\bf Acknowledgements}\\

\noindent
The work of A.G. was supported by the RFBR grant 18-52-05002 and the Tomsk Polytechnic University competitiveness enhancement program. I.M. acknowledges the support of
the RF Ministry of Education and Science under the project No 2.8172.2017/8.9.

\end{document}